\documentclass[letterpaper]{article} 
\usepackage{aaai25}  
\usepackage{times}  
\usepackage{helvet}  
\usepackage{courier}  
\usepackage[hyphens]{url}  
\usepackage{graphicx} 
\usepackage{natbib}  
\usepackage{caption} 
\usepackage{algorithm}
\usepackage{algorithmic}
\usepackage{multirow}
\usepackage{amsfonts}
\usepackage{array}
\usepackage{color}
\usepackage{pifont}
\usepackage{colortbl}
\usepackage{graphicx}
\usepackage{amsmath}
\usepackage{booktabs}
\usepackage{tabularx}
\usepackage{marvosym}
\newcommand{\cmark}{\ding{51}} 
\newcommand{\xmark}{\ding{55}} 
\newcolumntype{Y}{>{\centering\arraybackslash}X}
\usepackage{newfloat}
\usepackage{listings}
\DeclareCaptionStyle{ruled}{labelfont=normalfont,labelsep=colon,strut=off} 
\floatstyle{ruled}
\newfloat{listing}{tb}{lst}{}
\floatname{listing}{Listing}
\title{Thinking in Granularity: Dynamic Quantization for Image Super-Resolution by Intriguing Multi-Granularity Clues}
\author{
    Mingshen Wang\textsuperscript{\rm 1},
    Zhao Zhang\textsuperscript{\rm 1,\rm 2\Letter},
    Feng Li\textsuperscript{\rm 1\Letter},
    Ke Xu\textsuperscript{\rm 3},
    Kang Miao\textsuperscript{\rm 1},
    Meng Wang\textsuperscript{\rm 1}
}
\affiliations{
    \textsuperscript{\rm 1}School of Computer Science and Information Engineering, Hefei University of Technology, Hefei, China\\
    \textsuperscript{\rm 2}Yunnan Key Laboratory of Software Engineering, Yunan, China\\
    \textsuperscript{\rm 3}School of Artificial Intelligence, Anhui University, Hefei, China\\
}

\usepackage{bibentry}

\begin{document}
\maketitle
\begin{abstract}
Dynamic quantization has attracted rising attention in image super-resolution (SR) as it expands the potential of heavy SR models onto mobile devices while preserving competitive performance. Existing methods explore layer-to-bit configuration upon varying local regions, adaptively allocating the bit to each layer and patch. Despite the benefits, they still fall short in the trade-off of SR accuracy and quantization efficiency. Apart from this, adapting the quantization level for each layer individually can disturb the original inter-layer relationships, thus diminishing the representation capability of quantized models. In this work, we propose Granular-DQ, which capitalizes on the intrinsic characteristics of images while dispensing with the previous consideration for layer sensitivity in quantization. Granular-DQ conducts a multi-granularity analysis of local patches with further exploration of their information densities, achieving a distinctive patch-wise and layer-invariant dynamic quantization paradigm. Specifically, Granular-DQ initiates by developing a granularity-bit controller (GBC) to apprehend the coarse-to-fine granular representations of different patches, matching their proportional contribution to the entire image to determine the proper bit-width allocation. On this premise, we investigate the relation between bit-width and information density, devising an entropy-to-bit (E2B) mechanism that enables further fine-grained dynamic bit adaption of high-bit patches. Extensive experiments validate the superiority and generalization ability of Granular-DQ over recent state-of-the-art methods on various SR models. Code and supplementary statement can be found at \url{https://github.com/MmmingS/Granular-DQ.git}.
\end{abstract}
%
\begin{figure}[t]
	\centering
	\includegraphics[width=\linewidth]{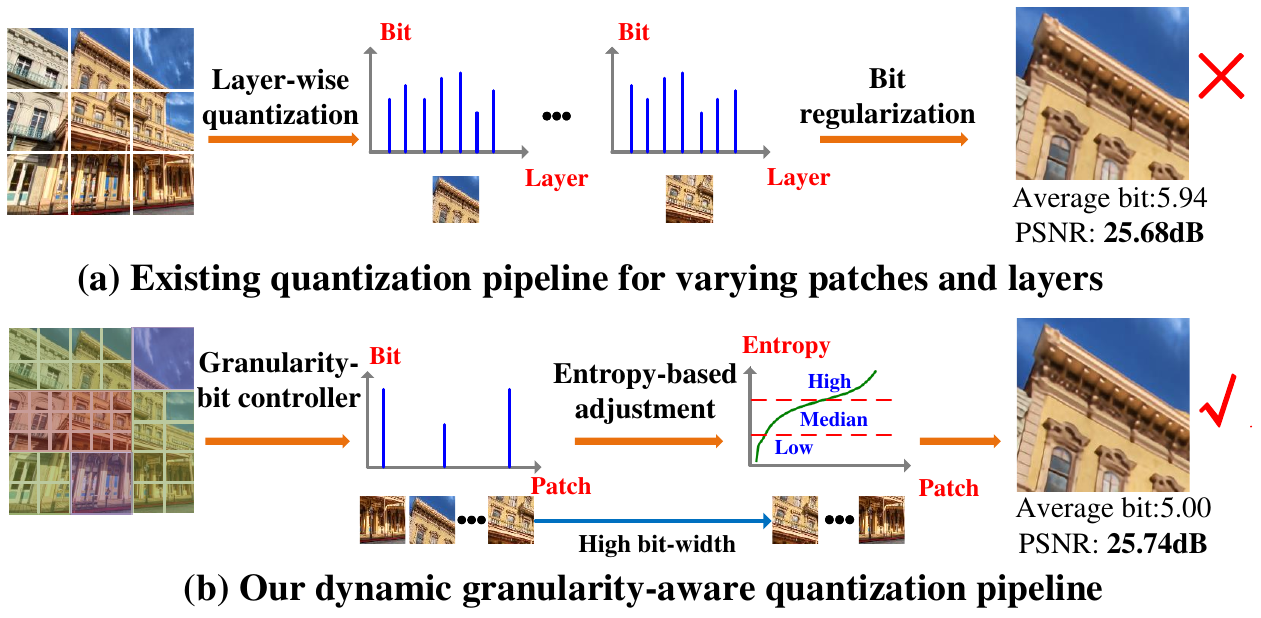}
	\caption{Visual comparison of (a) previous dynamic quantization pipeline~\cite{hong2022cadyq} that adapt the bit allocation for layers and patches simultaneously and (b) our Granular-DQ pipeline conducts patch-wise and layer-invariant dynamic quantization, which contains two steps: 1) granularity-aware bit allocation and 2) fine-grained bit-width adaption based on the entropy statistics. Our method recovers a better SR image with a lower average bit.}
    \vspace{-0.1cm}
	\label{fig:teaser}
    \vspace{-0.2cm}
\end{figure}

\section{Introduction}
\label{sec:introduction}
Single image super-resolution (SISR) has been a fundamental task in the computer vision community, aiming to recover high-resolution (HR) images from corrupted low-resolution (LR) input. Recently, from the pioneering deep learning-based method~\cite{dong2014learning}, convolutional neural networks (CNN)~\cite{dong2014learning,kim2016accurate,shi2016real,zhang2018image,ahn2018fast,li2022dssr} and transformers~\cite{liang2021swinir,lu2022esrt,zhang2022elan,chen2023hat} have dominated SISR. While the SR performance continues to achieve breakthroughs, the model complexity of later methods also increases constantly, which limits their practical applications, especially tackling large-size images (\emph{e.g.} 2K and 4K). This raises interest in compressing deep SR models to unlock their potential on resource-constrained devices. 

Model quantization~\cite{zhou2016dorefa} has emerged as a promising technology that reduces both computational overhead and memory cost with minimal performance sacrifice, where the effectiveness has been demonstrated in a wide range of high-level tasks~\cite{zhou2016dorefa,choi2018pact,bhalgat2020lsq+,Chen2021cvpr,gao2022towards,luo2023long}. Some prior works design SR quantizers by adjusting the quantization range~\cite{li2020pams,zhong2022dynamic} or modeling the feature distribution ~\cite{Hong_2022_WACV,qin2024quantsr} for activations, assigning a fixed bit for diverse image regions. However, these methods overlook that the accuracy degradation from quantization can vary for different contents, where some are more sensitive to quantization, thus showing a worse tolerance for low bits.

To address this limitation, Hong~\emph{et al.}~\cite{hong2022cadyq} propose content-aware dynamic quantization (CADyQ) which employs trainable bit selectors to measure the image and layer sensitivities for quantization simultaneously, as illustrated in Figure~\ref{fig:teaser}(a). Nevertheless, incorporating such selectors into each layer will cause additional computational costs, particularly pronounced in deep networks. Several methods~\cite{Tian_2023_CVPR,lee2024refqsr} improve the trained selectors in CADyQ by exploring different image characteristics of patches, which conduct once more patch-wise quantization to tackle the image sensitivity. Though some advancements have been made, such a layer-wise bit-width adaption in response to varying patches can introduce disturbances to the inter-layer relations within original models to some extent, which leads to disparities in the representations, consequently compromising the reconstruction after quantization.

These observations prompt us to consider a key question: \emph{Can we straightly adapt quantization with the awareness of image contents while avoiding layer sensitivity?} In this context, deviating from existing methods, we rethink the quantization principle from two perspectives: 1) Granular characteristic, where fine-granularity representations reveal the texture complexity of local regions and coarse ones express structural semantics of the overall scene; 2) Entropy statistic, which reflects the average information density and the complexity of pixel distributions given patches~\cite{shannon}, correlated with the image quality. Therefore, we propose a distinctive approach, dubbed Granular-DQ, which conducts low-bit dynamic quantization by harnessing the multi-granularity clues of diverse image contents to achieve efficient yet effective quantized SR models.

Granular-DQ consists of two sequential policies: one to conduct granularity-aware bit allocation for all the patches and the other is fine-grained bit-width adaption based on the entropy (see Figure~\ref{fig:teaser}(b)). For the former, we design a granularity-bit controller (GBC) that constructs a hierarchy of coarse-to-fine granularity representations for each patch. GBC then assigns an appropriate level of granularity to each patch, contingent upon its desired contribution percentage to the entire image, and aligns this with potential quantization bit-widths, enabling a tailored bit allocation. However, since Granular-DQ contains no bit constraint as CADyQ, relying solely on the GBC for quantization will force the network to be optimized toward reconstruction accuracy with pixel-wise supervision, leading to excessively high bits on some patches. To alleviate this, we present an entropy-based fine-tuning approach on the premise of GBC, making a fine-grained bit adjustment for the patches less quantized. We capture generalized distribution statistics of the entropy across large-scale data, providing approximate entropy thresholds to establish an entropy-to-bit (E2B) mechanism. The resultant entropy thresholds are then dynamically calibrated and fine-tuned by exploiting the entropy of calibration patches as the adaption factor, achieving a more precise bit assignment. Experiments on representative CNN- and transformer-based SR models demonstrate the superiority of Granular-DQ in the trade-off between accuracy and quantization efficiency over recent state-of-the-art methods. 
The main contributions are summarized as follows:

\begin{itemize}
	\item For the first time, we propose Granular-DQ, a markedly different method with full explorations of the granularity and entropy statistic of images to quantization adaption, allowing complete patch-wise and layer-invariant dynamic quantization for SR models.
	\item We propose GBC which learns hierarchical granular representations of image patches and adaptively determines the granularity levels based on their contribution to the entire image, aligning these with suitable bit-widths.
    \item We propose an entropy-based fine-tuning approach upon GBC and build an E2B mechanism, which enables fine-grained and precise bit adaption for the patches with excessively high bits. Granular-DQ shows preferable performance with existing methods.
\end{itemize}

\section{Related Work}
\label{sec:related work}
\subsection{Single Image Super-Resolution.} 
Recent progress in CNNs has critically advanced the field of SISR, enhancing image quality and detail restoration significantly ~\cite{dong2014learning,lim2017enhanced}. 
However, the intensive computational demands of CNNs \cite{dong2014learning,shi2016real,zhang2018image,hui2018fast,li2019filternet}, transformer-based \cite{liang2021swinir,lu2022esrt,chen2023hat} and diffusion-based models \cite{rombach2022,sr32023} limit their use in mobile and embedded systems. Efforts to mitigate computational complexity have spanned several dimensions, research has focused on several strategies, including lightweight architecture implementation~\cite{chu2021fast,wang2021exploring}, knowledge distillation~\cite{hui2019lightweight,zhang2021data}, network pruning~\cite{zhang2021aligned}, re-parameterization~\cite{wang2022repsr}, and parameter sharing~\cite{chen2022arm}. Additionally, some adaptive networks have been investigated to refine both performance and efficiency dynamically~\cite{,chen2022arm,wang2022adaptive}, highlighting the ongoing pursuit of an optimal balance between resource occupation and SR performance. However, apart from the computational complexity, the obstacle of memory storage imposed by floating-point operations also limits the usage of existing SR models. This work applies the network quantization technique for this purpose.

\begin{figure*}[t]
	\centering
    \vspace{-0.1cm}
	\includegraphics[width=\linewidth]{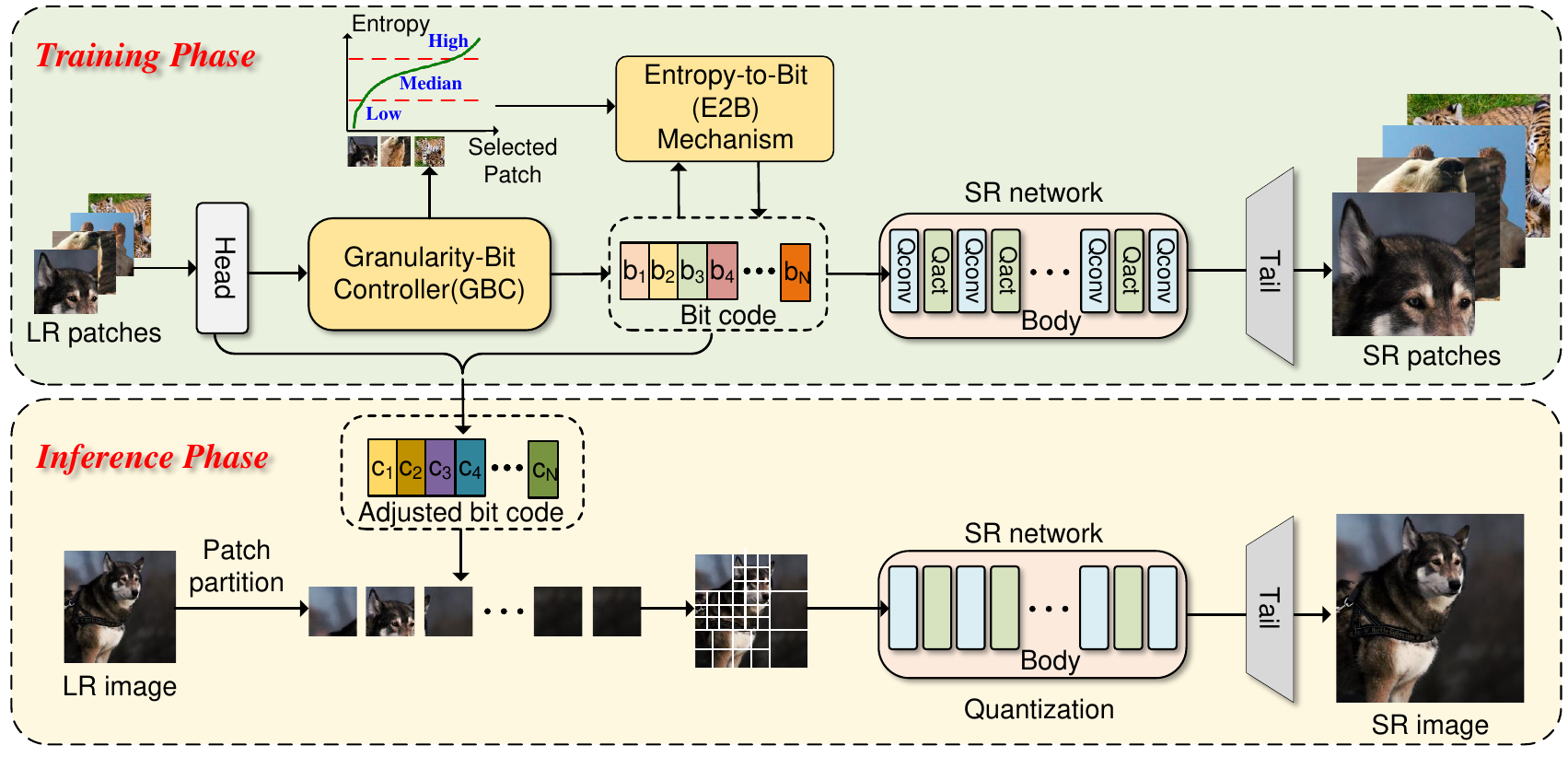}
	\caption{The schematic of the proposed Granular-DQ for SR networks. Granular-DQ is a patch-wise and layer-invariant quantization pipeline, which contains two key steps: 1) granularity-aware bit allocation by the granularity-bit controller (GBC) and 2) entropy-based fine-grained bit-width adaption on the patches allocated with high bits in GBC based on an entropy-to-bit (E2B) mechanism. During the inference phase, the input image is partitioned into serial patches mapped to the adapted bit code, which forces the SR network to be specifically quantized for each patch.}
    \vspace{-0.1cm}
	\label{fig:GDQ}
    \vspace{-0.2cm}
\end{figure*}


\subsection{Network Quantization} 
Network quantization has emerged as an effective solution that transforms 32-bit floating point values into lower bits~\cite{zhou2016dorefa,choi2018pact,zhuang2018towards,esser2019learned,bhalgat2020lsq+,li2021brecq} to improve the network efficiency, which can be divided into quantization-aware training (QAT) and post-training quantization (PTQ) methods. QAT~\cite{zhou2016dorefa,choi2018pact,esser2019learned,bhalgat2020lsq+} integrates the quantization process into the training of networks, performing quantization adaption with complete datasets. PTQ methods~\cite{li2021brecq,wei2022qdrop} often require a small calibration dataset to determine quantization parameters without retraining, which enables fast deployment on various devices. Recently, some methods introduce mixed-precision~\shortcite{dong2019hawq} or dynamic quantization~\shortcite{Liu2022dqnet} into the above two paradigms, which allows for the automatic selection of the quantization precision of each layer. Though network quantization has been predominantly applied in various high-level tasks, its potential in SISR has not been fully exploited.

\subsection{Quantization for Super-Resolution Networks}
Unlike high-level vision tasks, SISR presents unique challenges due to its high sensitivity to precision loss~\cite{li2020pams,wang2021fully,Hong_2022_WACV,hong2023overcoming}. PAMS~\cite{li2020pams} introduces the parameterized max scale scheme, which quantizes both weights and activations of the full-precision SR networks to fixed low-bit ones. DDTB~\cite{zhong2022dynamic} tackles the quantization of highly asymmetric activations by a layer-wise quantizer with dynamic upper and lower trainable bounds. DAQ~\cite{Hong_2022_WACV} and QuantSR~\cite{qin2024quantsr} study the influence of the parameter distribution in quantization, continuing to narrow the performance gap to full-precision networks. Recently, some attempts adopt dynamic quantization, which exploits the quantization sensitivity of layers and images, \emph{e.g.} gradient magnitude~\cite{hong2022cadyq}, edge score~\cite{Tian_2023_CVPR}, or cross-patch similarity~\cite{lee2024refqsr}, have demonstrated promising achievements. AdaBM~\cite{hong2024adabm} accelerates the adaptive quantization by separately processing image-wise and layer-wise bit-width adaption on the fly. In contrast, our method exploits the granularity and information density inherent in images to conduct dynamic quantization. It dispenses with the conventional need for layer sensitivity while being responsive to local contents, 
devising a distinctive patch-wise and layer-invariant dynamic quantization principle, which achieves superior performance and generalization ability for both CNN and transformer models.

\section{Proposed Method}
\label{sec:proposed method}
\subsection{Preliminaries}
\label{Preliminaries}
In most cases, converting the extensive floating-point calculations into operations that use fewer bits within CNNs involves quantizing the input features and weights at convolutional layers~\cite{krishnamoorthi2018quantizing}. 
In the quantized SR network, given a quantizer $\mathcal{Q}$ in a symmetric mode, the function $\mathcal{Q}_b(\cdot)$ is applied to the input $\hat{x}_k$ of the $k$-th convolutional layer, transforming $x_k$ into its quantized counterpart $\hat{x}_k$ with a lower bit-width $b$, as expressed in the following formula
\begin{equation}
	\hat{x}_k = \mathcal{Q}_b(x_k) = \operatorname{round}\left(\frac{\operatorname{clip}(\boldsymbol{x_k})}{r_b}\right)r_b,
	\label{eq:1}
\end{equation}
where $\operatorname{clip}(\cdot)=max(min(x_k,a),-a)$ confines $x_k$ within $[-a,a]$. $a$ denotes the maximum of the absolute value of $x$~\cite{wu2020integer} or derived from the moving average of max values across batches~\cite{wang2021fully}.
Additionally, $r_b$ serves as the mapping function that scales inputs of higher precision down to their lower bit equivalents, defined as $r_b = \frac{a}{2^{b-1}-1}$. Specially, the non-negative values after ReLU are truncated to $[0, a]$ and $r_b = \frac{a}{2^b-1}$. For weight quantization, given the $k$-th convolutional layer weight $w_k$, the quantized weight $\hat{w}_i$ can be formulated as follows
\begin{equation}
	\hat{w}_k = \mathcal{Q}_b(w_k) = \operatorname{round}\left(\frac{\operatorname{clip}(w_k)}{r_b}\right)r_b.
	\label{eq:2}
\end{equation}
Different from activations, the weights are quantized with fixed bit-width following~\cite{li2020pams,hong2022cadyq}.

\subsection{Granular-DQ for SISR}
\label{sec:gadq}
The proposed Granular-DQ aims to cultivate a layer-invariant SR quantization approach that enables dynamic quantization of existing SR models for varying image contents with the awareness of multi-granularity clues. The overall pipeline is shown in Figure~\ref{fig:GDQ}, which contains two steps: 1) granularity-aware bit allocation by the granularity-bit controller (GBC) and 2) entropy-based fine-grained bit-width adaption on the patches allocated with high bits in GBC based on an entropy-to-bit (E2B) mechanism. 

\begin{figure}[t]
    \centering
    \includegraphics[width=\linewidth]{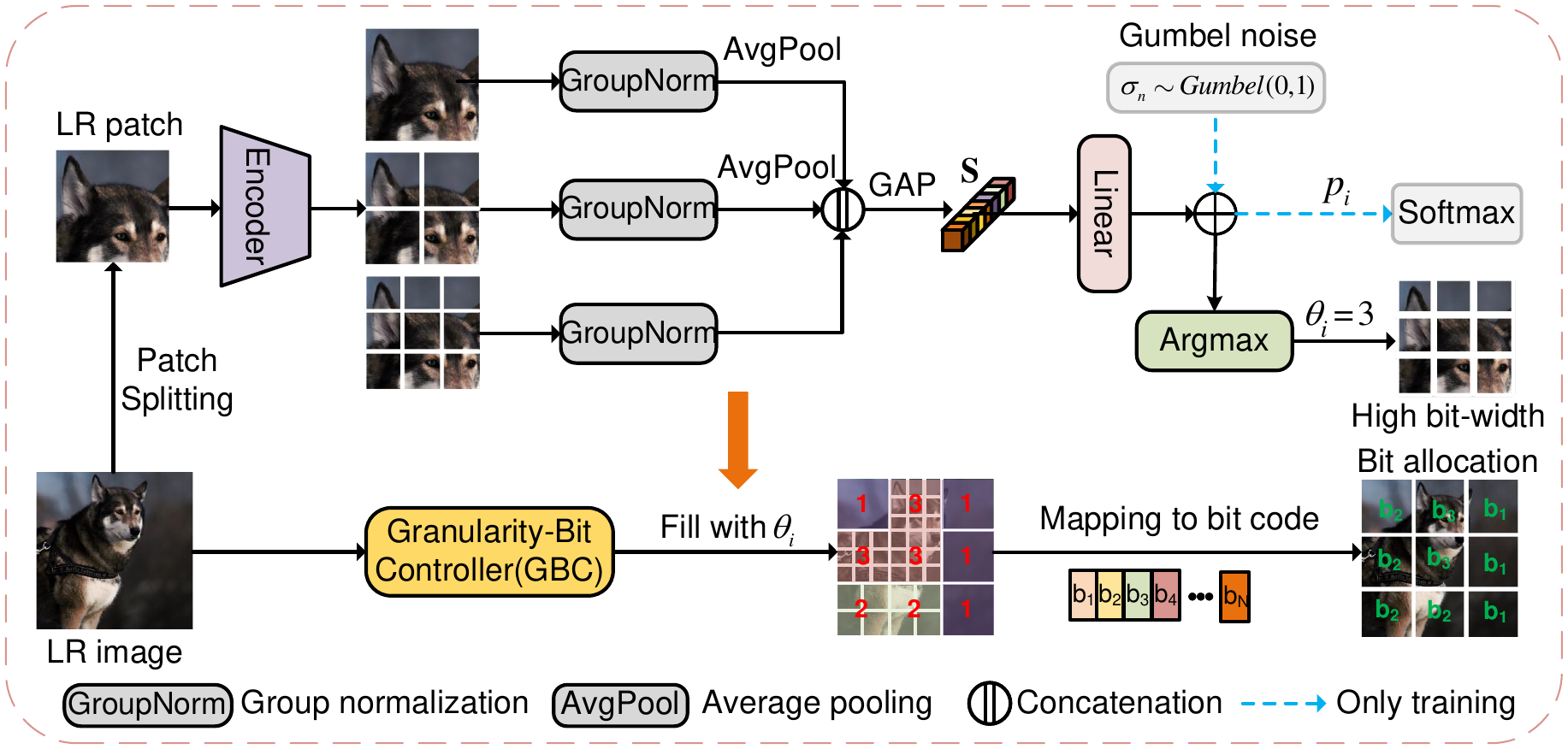}
    \caption{The structure of granularity-bit controller (GBC). It constructs hierarchical coarse-to-fine granularity representations for each patch. Then, it measures the granularity level of the patch upon its desired contribution percentage to the entire image, and maps this to quantization bit codes, finally achieving a tailored bit allocation.}
    \vspace{-0.1cm}
    \label{fig:DGAM}
    \vspace{-0.2cm}
\end{figure}

\noindent\textbf{Granularity-Bit Controller}. Given an image $X$, as shown in Figure~\ref{fig:DGAM}, the GBC first encodes it into hierarchical feature $\mathbf{Z}=\mathcal{E}(X)$ by the encoder $\mathcal{E}$, where $\mathbf{Z}={Z_1, Z_2, ..., Z_D}$ via $D-1$ downsampling operations. Note that the resolution from $Z_1$ to $Z_D$ decreases progressively, where the largest $Z_1$ corresponds to the finest-granularity feature and the smallest $Z_D$ denotes the coarsest-granularity one (\emph{i.e.} $D$ granularities), forming multi-granularity representations for $X$. We implement GBC with the Gumbel-Softmax, a differentiable sampling scheme~\cite{jang2016categorical}, to adaptively measure the proportional contribution of all patches to the entire image, and align this with potential quantization bit-widths. To be specific, all the granularity features are group normalized and then average pooled to the coarsest granularity, \emph{i.e.}, with the same resolution of $Z_D$, denoted by $\mathbf{\hat{Z}}={\hat{Z}_1, \hat{Z}_2, ..., \hat{Z}_D}$. We concatenate $\mathbf{\hat{Z}}$ along the channel dimension and squeeze the multi-granularity information by global average pooling $GAP(\cdot)$ to generate a channel-wise statistics $\mathbf{S}$ of $X$, formulated by
\begin{equation}
    \mathbf{S}=GAP(\Vert\hat{Z}_1, \hat{Z}_2, ..., \hat{Z}_D\Vert).
\end{equation}

Assuming there are $N$ total bit codes ($b_1,...,b_n,...,b_N$) with different bit-widths, a linear layer is employed to acquire a learnable weight $\mathbf{W}_\mathbf{g}\in\mathbb{R}^{(N\times D)\times N}$ that operates on $\mathbf{S}$ to generate the gating logits $\mathbf{G}\in\mathbb{R}^{1\times1\times N}$ as
\begin{equation}
    \mathbf{G}=\mathbf{W}_\mathbf{g}\mathbf{S},
	\label{eq:4}
\end{equation}
For each patch $X_i$, its gating logit $g_i\in\mathbb{R}^N$ is utilized to ascertain the granularity level through the gating index $\theta_i$:
\begin{equation}
	\theta_i=\arg\max_n(g_{i, n})\in\{1,2,..., n\}.
	\label{eq:5}
\end{equation}

\begin{figure}[t]
    \centering
    \includegraphics[width=0.85\linewidth]{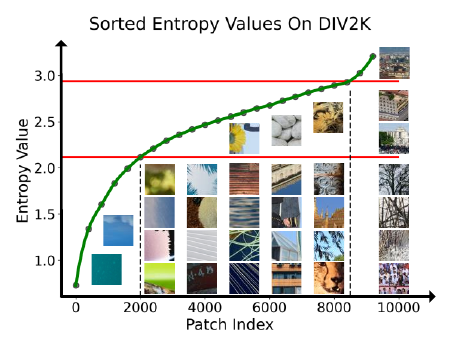}
    \caption{The generalized distribution statistic of the entropy for all LR patches on DIV2K.}
    \vspace{-0.1cm}
    \label{fig:entropy}
    \vspace{-0.2cm}
\end{figure}

Inspired by the end-to-end discrete methodology in \cite{xie2020spatially}, the fixed decision typically dictated by Eq.(\ref{eq:5}) is substituted with a probabilistic sampling approach. It hinges on the utilization of a categorical distribution characterized by unnormalized log probabilities, from which discrete gating indices are derived by integrating a noise sample $\sigma_{n}$, originating from the standard Gumbel distribution $\mathrm{Gumbel}(0,1)$:
\begin{equation}	\theta_i=\arg\max_n(g_{i,n}+\sigma_n).
	\label{eq:6}
\end{equation}
After that, we calculate the gating score $p_i$ for each patch:
\begin{equation}
p_i=\frac{\exp((g_{i,\theta_i}+\sigma_{\theta_i}))/\tau}{\sum_n^N\exp((g_{i,n}+\sigma_n)/\tau)},
	\label{eq:7}
\end{equation}
where $p_i\in[0,1]$ measures the probability of $X_i$ contributing to the entire image $X$, thus determining the granularity level and pointing to a corresponding code $b_n$. In our experiments, we set the temperature coefficient $\tau=1$. Similar to the forward propagation approach in quantization, the gradients for such a gate are calculated using a straight-through estimator, derived from $p_i$ during the backward pass. By incorporating GBC at the onset of SR networks, Granular-DQ only introduces negligible computational overhead.

\begin{table*}[t]
\footnotesize
    \setlength\tabcolsep{3pt}
    \setlength{\abovecaptionskip}{5pt}
    \centering
    \small
    \vspace{-0.1cm}
    \begin{tabularx}{\textwidth}{@{\extracolsep{\fill}}llYYYYYYYYY}
        \toprule
        \multirow{2}{*}{\centering Methods} & \multirow{2}{*}{\centering Scale} & \multicolumn{3}{c}{Urban100} & \multicolumn{3}{c}{Test2K} & \multicolumn{3}{c}{Test4K} \\ 
        \cmidrule(lr){3-5} \cmidrule(lr){6-8} \cmidrule(lr){9-11} 
         & & FAB↓ & PSNR↑ & SSIM↑ & FAB↓ & PSNR↑ & SSIM↑ & FAB↓ & PSNR↑ & SSIM↑ \\
        \midrule
        SRResNet & $\times4$ & 32.00 & 26.11  & 0.787 & 32.00 & 27.65  & 0.776  & 32.00 & 29.04  & 0.823  \\
        PAMS  & $\times4$ & 8.00  & 26.01  & 0.784 & 8.00  & 27.67  & 0.781  & 8.00  & 28.77  & 0.813   \\
        CADyQ & $\times4$ & 5.73  & 25.92  & 0.781 & 5.14  & 27.64  & 0.781  & 5.02  & 28.72  & 0.812   \\ 
        CABM  & $\times4$ & 5.34  & 25.86  & 0.778 & 5.17  & 27.52  & 0.771  & 5.07  & 28.91  & 0.818   \\ 
        AdaBM & $\times4$ & 5.60  & 25.72  & 0.773 & 5.20  & 27.55  & 0.777  & 5.10  & 28.62  & 0.809   \\
        RefQSR($\delta$\text{-4bit}) & $\times4$ & 4.00  & 25.90  & 0.778  & 5.17  & 27.52 & 0.771 & 5.07  & 28.91  & 0.818   \\
        Granular-DQ (Ours) & $\times4$ & \textbf{4.00}  & \textbf{25.98}  & \textbf{0.783}  & \textbf{4.01}  & \textbf{27.55}  & \textbf{0.773} & \textbf{4.01}  & \textbf{28.93}  & \textbf{0.820}   \\ 
        \midrule
        EDSR  & $\times4$ & 32.00 & 26.03  & 0.784 & 32.00 & 27.59  & 0.773 & 32.00 & 28.80  & 0.814  \\
        PAMS  & $\times4$ & 8.00  & 26.01  & 0.784 & 8.00  & 27.67  & 0.781 & 8.00  & 28.77  & 0.813   \\
        CADyQ & $\times4$ & 6.09  & 25.94  & 0.782 & 5.52  & 27.67  & 0.781 & 5.37  & 28.91  & 0.818   \\
        CABM  & $\times4$ & 5.80  & 25.95  & 0.782 & 5.65  & 27.57  & 0.772 & 5.56  & 28.96  & 0.819   \\
        Granular-DQ (Ours) & $\times4$ & \textbf{4.97}  & \textbf{26.01}  & \textbf{0.784} & \textbf{4.57}  & \textbf{27.58}  & \textbf{0.773} & \textbf{4.41}  & \textbf{28.98}  & \textbf{0.820}   \\
        \midrule
        IDN   & $\times4$ & 32.00 & 25.42  & 0.763  & 32.00 & 27.48  & 0.774 & 32.00 & 28.54  & 0.806  \\
        PAMS  & $\times4$ & 8.00  & 25.56  & 0.768  & 8.00  & 27.53  & 0.775 & 8.00  & 28.59  & 0.807   \\
        CADyQ & $\times4$ & 5.78  & 25.65  & 0.771  & 5.16  & 27.54  & 0.776 & 5.03  & 28.61  & 0.808   \\ 
        CABM  & $\times4$ & 4.28  & 25.57  & 0.768  & 4.25  & 27.42  & 0.766 & 4.23  & 28.74  & 0.813   \\ 
        Granular-DQ (Ours) & $\times4$ & \textbf{4.18}  & \textbf{25.68}  & \textbf{0.772} & \textbf{4.29}  & \textbf{27.47}  & \textbf{0.767} & \textbf{4.23}  & \textbf{28.83}  & \textbf{0.816}   \\  
        \midrule 
        SwinIR-light & $\times4$ & 32.00 & 26.46  & 0.798  & 32.00 & 27.72  & 0.779 & 32.00 & 29.14  & 0.825  \\
        PAMS      & $\times4$ & 8.00  & 26.31  & 0.793  & 8.00  & 27.67  & 0.776 & 8.00  & 29.08  & 0.823   \\
        CADyQ     & $\times4$ & 5.15  & 25.87  & 0.779  & 5.01  & 27.54  & 0.772 & 5.01  & 28.92  & 0.819  \\ 
        CABM      & $\times4$ & 5.34  & 25.88  & 0.780  & 4.92  & 27.62  & 0.774 & 4.91  & 29.02  & 0.821   \\  
        Granular-DQ (Ours)  & $\times4$ & \textbf{4.79}  & \textbf{26.42}  & \textbf{0.796}  & \textbf{4.74} & \textbf{27.67}  & \textbf{0.778}  & \textbf{4.76}  & \textbf{29.11}  & \textbf{0.824}   \\ 
        \midrule 
        HAT-S    & $\times4$ & 32.00 & 27.81  & 0.833  & 32.00 & 28.07  & 0.791  & 32.00 & 29.56 & 0.836   \\
        PAMS      & $\times4$ & 8.00  & 27.56  & 0.827  & 8.00  & 28.00  & 0.789  & 8.00  & 29.48 & 0.834   \\
        CADyQ     & $\times4$ & 5.53  & 26.98  & 0.814  & 5.41  & 27.88  & 0.784  & 5.33  & 29.32 & 0.830  \\ 
        CABM      & $\times4$ & 5.49  & 26.95  & 0.813  & 5.38  & 27.87  & 0.784  & 5.30  & 29.31 & 0.829  \\  
        Granular-DQ (Ours)  & $\times4$ & \textbf{4.77}  & \textbf{27.66}  & \textbf{0.829}  & \textbf{4.80} & \textbf{28.01}  & \textbf{0.789}  & \textbf{4.78}  & \textbf{29.49}  & \textbf{0.834}   \\
        \midrule
    \end{tabularx}
    \caption{Quantitative comparison (FAB, PSNR (dB)/SSIM) with full precision models, PAMS, CADyQ, CABM, RefQSR and our method on Urban100, Test2K, Test4K for $\times 4$ SR. $\times 2$ SR results are provided in the \textbf{supplementary material}.}
    \vspace{-0.1cm}
    \label{tab_1}
    \vspace{-0.2cm}
\end{table*}

\noindent\textbf{Entropy-based Fine-grained Bit-width Adaption}. In this work, since Granular-DQ is optimized by pixel-wise supervision, relying solely on the GBC for quantization will force the network to be optimized toward reconstruction accuracy with pixel-wise supervision, which can lead to excessively high bits on some patches. To tackle this problem, we propose an entropy-based scheme to fine-tune bit adaption on the patches less quantized by GBC.

Specifically, we capture a generalized distribution statistic of the entropy for all LR patches on the training set. We first discretize the total $N$ pixels within a patch into multiple bin intervals $B$ based on the pixel values, which can estimate the probability distribution of pixels smoothly. Then entropy is computed as
\begin{equation}
	\mathcal{H}=-\sum_{i=1}^{N}\mathcal{P}(x_i)log(\mathcal{P}(x_i)).
    \label{eq:8}
    \vspace{-0.1cm}
\end{equation}
We use Gaussian-weighted kernel to assign different importance to the pixels in a patch with the formulation of ${\sum_{i=1}^N\sum_{j=1}^Bexp(-\frac{(r_i)^2}{2\sigma^2})+\epsilon}$, where $r_i$ denotes the residual between the pixel value of the $i$-th pixel $x_i$ and the segment values for bin intervals. Thus, one can obtain its kernel density $\mathcal{P}(x_i)$ by $\frac{\sum_{j=1}^Bexp(-\frac{(r_i)^2}{2\sigma^2})}{\sum_{i=1}^N\sum_{j=1}^Bexp(-\frac{(r_i)^2}{2\sigma^2})+\epsilon}$. In this way, we can get the entropy statistic across the overall training set, represented by $\mathbf{H}={\mathcal{H}_1,\mathcal{H}_2,...,\mathcal{H}_M}$ sorted in ascending order with $M$ patches, as shown in Figure~\ref{fig:entropy}.


We establish an entropy-to-bit (E2B) mechanism based on the entropy statistic $\mathbf{H}$ and conduct fine-grained bit-width adjustment. Firstly, serial quantiles are inserted on $\mathbf{H}$ to divide it into multiple subintervals $V$ by $\mathcal{I}_t = \lceil\frac{M\cdot t}{V}\rceil$, where $\mathcal{I}_t$ denotes the patch indice at the $t$-th quantile, which points to a certain entropy $\mathcal{H}_t$ in $\mathbf{H}$. The quantiles can be seen as thresholds, thus we provide candidate bit configurations according to the thresholds for all the patches. Given a patch with its entropy $E$, one can find the index of the subinterval in $\mathbf{H}$, and finally determine the adapted bit-width. Taking two quantiles $t_1$ and $t_2$ as an example, we can get two patch indices $\mathcal{I}_{t_1}$ and $\mathcal{I}_{t_2}$ which corresponds to the entropy values $\mathcal{H}_{t_1}$ and $\mathcal{H}_{t_2}$ respectively, \emph{i.e.} $\mathbf{H}$ will be divided into three discrete subintervals as 
\begin{equation}
	c_n=\begin{cases}c_1&\text{if }E\leq \mathcal{H}_{t_1},\\
    c_2&\text{if }\mathcal{H}_{t_1}<E\leq \mathcal{H}_{t_2},\\
    c_3&\text{if }\mathcal{H}_{t_2}<E\leq \mathcal{H}_M\end{cases}
	\label{eq:9}
\end{equation}
where $c_n$ denotes the adapted bit codes.

To further improve the flexibility and robustness of E2B for various contents, we present an adaptive threshold calibration (ATC) scheme on E2B. During the training iterations $J$, we leverage the exponential moving average (EMA) to dynamically calibrate the threshold $t$, formulated by
\begin{equation}
t^{(j)}=t^{(j-1)} \cdot \gamma + {Norm(E)} \cdot (1 - \gamma),
\label{eq:10}
\end{equation}
where $Norm(\cdot)=\frac{\mathcal{H}_t - \mathcal{H}_{min}}{\mathcal{H}_{max} - \mathcal{H}_{min}}$, and $\mathcal{H}_{max}$ and $\mathcal{H}_{min}$ denotes the maximum and minimum entropy of all the patches in the current mini-batch at the $j$-th iteration. $\gamma$ represents the smoothing parameter of EMA, which is set to 0.9997. It should be noted that the LR samples remain consistent across epochs during training. Hence, our method only necessitates the E2B with ATC at the initial epoch, circumventing significant computational expenditure with iterations. Once the model is trained, as shown in Figure~\ref{fig:GDQ}, our method enables to fine-grained adapt the bit-widths of the patches based on calibrated thresholds from the large training set, yielding preferable bit codes $[c_1,c_2,...,c_N]$.  

In summary, by combining GBC and E2B, our method ensures optimal bit allocation for each patch individually while dispensing with the consideration for layer sensitivity as previous methods~\cite{hong2022cadyq,Tian_2023_CVPR}.

\begin{figure*}[t]
	\centering
    \vspace{-0.1cm}
	\includegraphics[width=\linewidth]{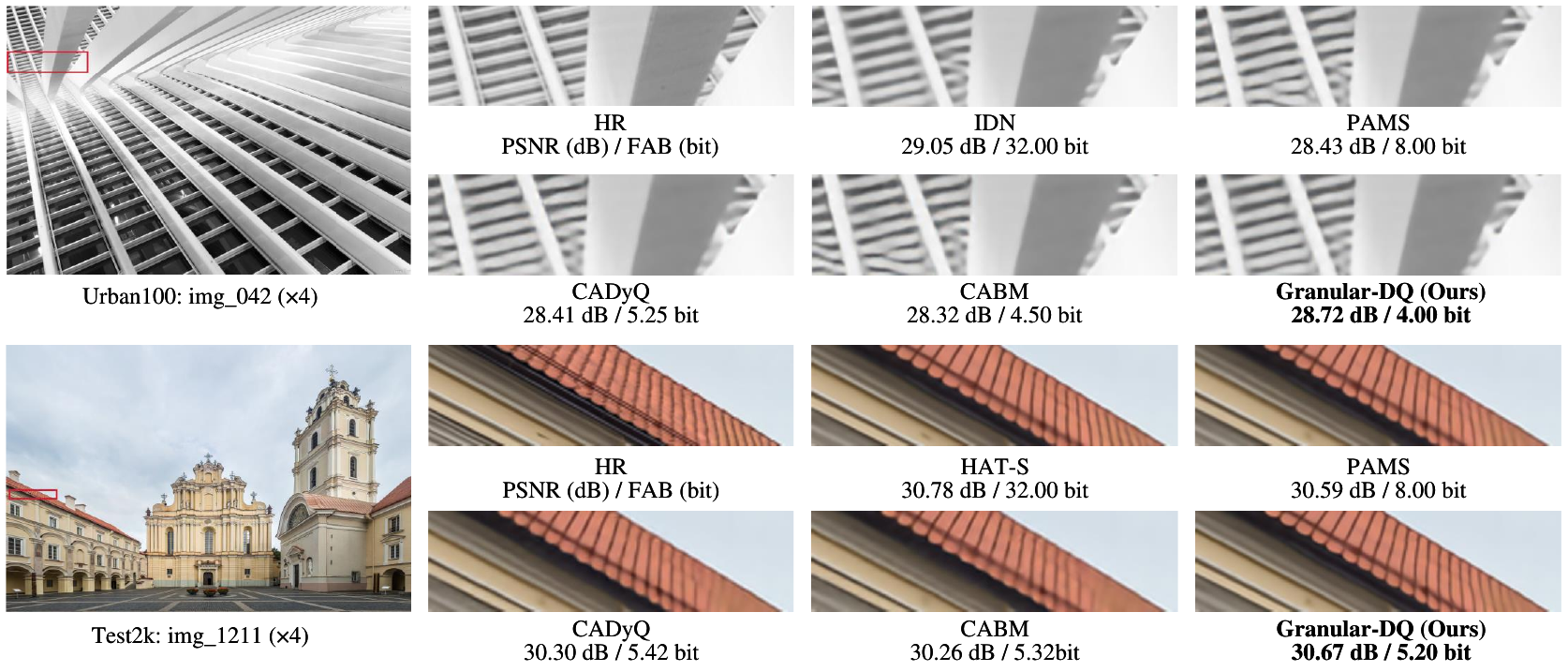}
	\caption{Qualitative comparison ($\times$4) on Urban100 and Test2K based on IDN and HAT-S models. Granular-DQ reconstructs SR images with better details and quantitative results}
	\vspace{-0.2cm}
    \label{fig:sr}
    \vspace{-0.15cm}
\end{figure*}

\subsection{Loss Function}
In previous SR quantization methods~\cite{hong2022cadyq,Tian_2023_CVPR,lee2024refqsr}, the objective function is composed of L1 loss, knowledge distillation loss, and even bit regularization term to facilitate the bit adaption. In Granular-DQ, we only use $L_1$ loss to train all the models
\begin{equation}
    L_1=\left\|I_{HR}-I_{SR}\right\|_1
	\label{eq:11}
\end{equation}
where $I_{HR}$ is the HR ground truth of the LR input and $I_{SR}$ is the SR reconstruction by our Granular-DQ.

\section{Experiments}
\label{experiments}
\subsection{Experimental Settings}
\textbf{Baseline SR Models.} The proposed Granular-DQ is applied directly to existing CNN-based SR models including SRResNet~\cite{ledig2017photo}, EDSR~\cite{lim2017enhanced}, and IDN~\shortcite{hui2018fast} as well as transformer-based models including SwinIR-light~\cite{liang2021swinir} and HAT-S~\cite{chen2023hat}. Following CADyQ~\cite{hong2022cadyq} and CABM~\cite{Tian_2023_CVPR}, we implement quantization on the weights and feature maps within the high-level feature extraction part, which is the focal point for the majority of computationally intensive operations. Notably, for SwinIR-light and HAT-S, the attention blocks are computed with full precision due to severe quantization errors, where more details are provided in the \textbf{supplementary material}. 

In Granular-DQ, the first step for bit allocation by GBC designates 4/6/8-bit as the candidate bits to quantize the patches. Subsequently, the second step by E2B adapts the patches allocated with 8 bits in GBC are further adapted using 4/5/8-bit as the candidates for fine-grained bit-width adjustment.  The initial entropy thresholds, denoted as $t_1$ and $t_2$, are set to 0.5 and 0.9 respectively and then gradually calibrated according to the entropy statistic on the training set, for all models. In this work, we employ QuantSR~\cite{qin2024quantsr} for all the quantization candidates and uniformly apply 8-bit linear quantization for weights.

\noindent\textbf{Datasets and Metrics.} In our experiments, all the models are trained on DIV2K~\cite{agustsson2017ntire} dataset which contains 800 training samples for $\times 2$ and $\times 4$ SR. We evaluate the model and compare it with existing methods on three benchmarks: Urban100~\cite{huang2015single}, Test2K and Test4K~\cite{kong2021classsr} derived from DIV8K dataset~\cite{gu2019div8k} by bicubic downsampling. We quantitatively measure the SR performance 
 using two metrics: peak signal-to-noise ratio (PSNR) and the structural similarity index (SSIM) for reconstruction accuracy. Besides, we also compute the 
 feature average bit-width (FAB) which represents the average bit-width across all features within the test dataset to measure the quantization efficiency.

 \noindent\textbf{Implementation details.} During training, we randomly crop each LR RGB image into a $48\times 48$ patch with a batch size of 16. All the models are trained for 300K iterations on NVIDIA RTX 4090 GPUs with Pytorch. The learning rate is set to $2\times 10^{-4}$ and is
 halved after 250K iterations. During testing, the input image is split into $96\times 96$ LR patches.

 \subsection{Comparing with the State-of-the-Art}
\textbf{Quantitative Comparison.} Table~~\ref{tab_1} reports the quantitative results on benchmarks. The proposed Granular-DQ is compared with original full-precision models, PAMS~\cite{li2020pams}, CADyQ, CABM, AdaBM~\cite{hong2024adabm}, and  RefQSR~\shortcite{lee2024refqsr}. One can see that Granular-DQ demonstrates the minimum performance sacrifice relative to the full-precision SRResNet and EDSR models while attaining the lowest FAB against other methods on all benchmarks. For IDN, Granular-DQ even exceeds its full-precision model by about 0.2dB on Urban100 and Test4K datasets, whereas other methods show lower PSNR and SSIM improvements with obviously higher FAB. Moreover, when implementing these methods on transformer-based baselines, it can be observed that Granular-DQ significantly outperforms other methods in terms of reconstruction accuracy and quantization efficiency. The results validate the superior effectiveness and generalization ability of Granular-DQ.

\noindent\textbf{Qualitative Comparison.}
Figure~\ref{fig:sr} shows the qualitative results on the Urban100 dataset. As one can see, Granular-DQ produces SR images with sharper edges and clearer details, sometimes even better than the original unquantized IDN. By comparison, despite the lower PSNR and more FAB consumption, existing methods also suffer from obvious blurs and misleading textures. 

\begin{table}[t]
\footnotesize
\centering
\setlength\tabcolsep{4pt}
    \begin{tabularx}{\hsize}{@{\extracolsep{\fill}}lYcr}
        \toprule
        \multirow{2}{*}{Method} & \multirow{2}{*}{FAB } & \multicolumn{1}{c}{Params (K)} & \multicolumn{1}{c}{BitOPs (G)} \\ 
        & & (↓ Ratio) & \multicolumn{1}{c}{(↓ Ratio)} \\
        \midrule
            EDSR  & 32.00  & 1518K (0.0\%) & 527.0T (0.0\%)     \\ 
        \midrule
            PAMS  & 8.00   & 631K (↓ 58.4\%) & 101.9T (↓ 80.7\%)     \\ 
            CADyQ & 6.09   & 489K (↓ 67.8\%) & 82.6T (↓ 84.3\%)    \\ 
            CABM  & 5.80   & \textbf{486K (↓ 68.0\%)} & 82.4T (↓ 84.4\%)     \\
            Ours & \textbf{4.97}  & \textbf{486K (↓ 68.0\%)} & \textbf{73.6T (↓ 86.0\%)}   \\ 
        \hline   
\end{tabularx}
\caption{Model complexity and compression ratio of EDSR for different quantization methods. We calculate the average BitOPs for generating SR images on the Urban100 dataset.}
\vspace{-0.1cm}
\label{tab_2}
\vspace{-0.2cm}
\end{table}

\noindent\textbf{Complexity Analysis.} To further investigate the complexity of our method for quantizing SR models, we calculate the number of operations weighted by the bit-widths (BitOPs)~\cite{van2020bayesian} as the metric and compare it with existing methods. As shown in Table~\ref{tab_2}, Granular-DQ leads to significant computational complexity reduction of the baseline model, which decreases the BitOPs from 527.0T to 73.6T and sustains a competitive FAB. Coupled with the decrease in the model parameters to 68.0\% (486K) of the full-precision model, the results demonstrate that Granular-DQ can ensure optimal trade-off between reconstruction accuracy and quantization efficiency. 

\subsection{Ablation Study}
\textbf{Effects of Individual Components.} We study the effects of the proposed components including GBC, E2B, and ATC in Table~\ref{tab_3}, where the results are evaluated on the Urban100 dataset. We can see that quantization with only GBC leads to a performance drop. Based on GBC, when we introduce E2B to conduct fine-grained bit-width adaption, the resultant quantizer can enhance the reconstruction accuracy and a small improvement in efficiency. Moreover, E2B and ATC in conjunction effectively reduce the FAB by a considerable margin (over 0.5 FAB) with almost the same PSNR/SSIM.

\begin{table}[t]
\footnotesize
\centering
\setlength\tabcolsep{3pt}
    \begin{tabularx}{\hsize}{@{\extracolsep{\fill}}YYYYYY}
        \toprule
        \multirow{2}{*}{GBC} & \multirow{2}{*}{E2B} & \multirow{2}{*}{ATC}  & \multicolumn{3}{c}{Urban100} \\
        \cmidrule(lr){4-6} 
         &  &  & FAB & PSNR & SSIM   \\ 
         \midrule
         \xmark & \xmark & \xmark &  8.00 & 26.01 & 0.783  \\ 
            \cmark & \xmark & \xmark & 5.86  & 25.97 & 0.782   \\ 
            \cmark & \cmark & \xmark & 5.51  & \textbf{26.02} & \textbf{0.784} \\ 
            \cmark & \cmark & \cmark & \textbf{4.97}  & 26.01 & \textbf{0.784}  \\ 
        \hline   
\end{tabularx}
\caption{Ablation study on individual proposed components in Granular-DQ including GBC, E2B, and ATC.}
\label{tab_3}
\end{table}

\begin{table}[t]
\footnotesize
\centering
\setlength\tabcolsep{3pt}
    \begin{tabularx}{\hsize}{@{\extracolsep{\fill}}YYYYYYY}
        \toprule
        \multirow{2}{*}{$b^{*}$} & \multicolumn{3}{c}{Set14} & \multicolumn{3}{c}{Urban100} \\
        \cmidrule(lr){2-4} \cmidrule(lr){5-7} 
         &  FAB & PSNR & SSIM &  FAB & PSNR & SSIM    \\ 
         \midrule
            $[4,5,6]$ & 5.29  & 28.52 & 0.780 & 4.85  & 25.98  & 0.783      \\ 
            $[4,5,7]$ & 5.50  & 28.54 & 0.780 & 4.98  & 25.99  & 0.782     \\
            $[4,6,7]$ & 5.79  & 28.55 & \textbf{0.781} & 5.22  & 25.97  & 0.783  \\
            $[4,6,8]$ & 5.64  & 28.57 &\textbf{0.781} & 5.38  & \textbf{26.01} & \textbf{0.784}      \\ 
            $[4,7,8]$ & 5.64  & 28.55 & 0.780 & 5.64  & 25.99  & 0.783      \\
            $[4,5,8]$ &\textbf{5.54} & \textbf{28.58} &\textbf{0.781} & \textbf{4.97}  & \textbf{26.01} & \textbf{0.784}     \\ 
        \hline   
\end{tabularx}
\caption{Ablation study on the influence of the bit configuration (denoted by $b^{*}$) in E2B with EDSR baseline.}
\label{tab_4}
\vspace{-0.3cm}
\end{table}


\noindent\textbf{Influence of the Candidate Bits in E2B.} We conduct experiments to investigate the influence of the bit configuration in E2B. For 3 candidate bits, we set the lowest bit-width as 4 and randomly change the other two, resulting in 6 variants. As reported in Table~\ref{tab_4}, the configuration of [4, 5, 6] performs worst on both Set14 and Urban100 with relatively lower FAB. Surprisingly, although we allocate higher bit-width to patches ([4, 7, 8]), the model incurs the most FAB but acquires negligible performance gains. By comparison, the model with [4, 5, 8] achieves the best trade-off on the two datasets, which is selected as our final configuration. \textbf{More ablations are provided in the supplementary material}. 


\section{Conclusion}
In this paper, we propose Granular-DQ, a patch-wise and layer-invariant approach that conducts low-bit dynamic quantization for SISR by harnessing the multi-granularity clues of diverse image contents. Granular-DQ constructs a hierarchy of coarse-to-fine granularity representations for each patch and performs granularity-aware bit allocation by a granularity-bit controller (GBC). Then, an entropy-to-bit (E2B) mechanism is introduced to fine-tune bit-width adaption for the patches with high bits in GBC. Extensive experiments indicate that our Granular-DQ outperforms recent state-of-the-art methods in both effectiveness and efficiency.

\section{Acknowledgments}
This work is supported by the National Natural Science Foundation of China (62472137, 62072151, 62302141, 62331003, 62206003), Anhui Provincial Natural Science Fund for the Distinguished Young Scholars (2008085J30), Open Foundation of Yunnan Key Laboratory of Software Engineering (2023SE103),  CCF-Baidu Open Fund (CCF-BAIDU202321), CAAI-Huawei MindSpore Open Fund (CAAIXSJLJJ-2022-057A) and the Fundamental Research Funds for the Central Universities (JZ2024HGTB0255).

{\small\bibliography{aaai25}}

\twocolumn[
    \centering
    {\LARGE\bf Supplementary Material\\
Thinking in Granularity: Dynamic Quantization for Image Super-Resolution by Intriguing Multi-   Granularity Clues \\}
    \vskip 0.1in
    {\Large \textbf{Mingshen Wang\textsuperscript{\rm 1},
    Zhao Zhang\textsuperscript{\rm 1,\rm 2*},
    Feng Li\textsuperscript{\rm 1*},
    Ke Xu\textsuperscript{\rm 3},
    Kang Miao\textsuperscript{\rm 1},
    Meng Wang\textsuperscript{\rm 1}} \\}
    \vskip 0.1in
    {\normalsize \textsuperscript{\rm 1}Hefei University of Technology\\
    \textsuperscript{\rm 2}Yunnan Key Laboratory of Software Engineering\\
    \textsuperscript{\rm 3}Anhui University
    }
]
\section{Appendix}


The supplementary material mainly includes the following contents:
\begin{itemize}
    \item The motivation of Granular-DQ.
    \item Ablation studies involve the thresholds (quantile $t$) number in E2B, and the combination with different weight quantization patterns, \emph{i.e.} PAMS~\cite{li2020pams} and QuantSR~\cite{qin2024quantsr}.
    \item More implementation details on the transformer-based baseline models including SwinIR-light~\cite{liang2021swinir} and HAT-S~\cite{chen2023hat}.
    \item More experimental results consist of $\times2$ SR performance and additional qualitative visualization.
    \item Limitations in Granular-DQ.
\end{itemize}

\subsection{Motivation}
Recent advances~\cite{Tu_2023_CVPR,hong2022cadyq,Tian_2023_CVPR,lee2024refqsr,hong2024adabm} have demonstrated the benefits of considering the quantization sensitivity of layers and image contents in SR quantization. Taking CADyQ~\cite{hong2022cadyq} for example, it applies a trainable bit selector to determine the proper bit-width and quantization level for each layer and a given local image patch based on the feature gradient magnitude. In our analysis, we compute the average bit-width and the quantization error measured by MSE between the reconstructions of the quantized model (via CADyQ) and the original high-precision model (EDSR) on the Test2K dataset. Figure~\ref{fig:motivation} (a) reveals that the majority of patches fall within a 6-bit to 8-bit range, accompanied by a relatively elevated MSE. Furthermore, we present t-SNE maps for various quantized layers and the final layer in Figure~\ref{fig:motivation} (c)-(d). Firstly, it is evident that the distribution of different layers quantized by CADyQ is markedly more scattered than that of the original model, as depicted in Figure~\ref{fig:motivation} (c). Secondly, on the final layer, the features from the CADyQ-quantized model exhibit a distinct vertical pattern, which is notably at odds with the structure of the original model's feature points (Figure~\ref{fig:motivation} (d)). In our investigation, CABM actually exhibited similar findings, although it fine-tunes the CADyQ model based on edge scores. These results indicate that: 1) Simply relying on image edge information is suboptimal for the trade-off between quantization efficiency and error; 2) The bit allocation for each layer in response to varying patches can introduce disturbances to the inter-layer relations within original models leading to disparities in the representations.

\renewcommand{\thefigure}{S\arabic{figure}}
\setcounter{figure}{0}  
\begin{figure}[t]
	\centering
	\includegraphics[width=\linewidth]{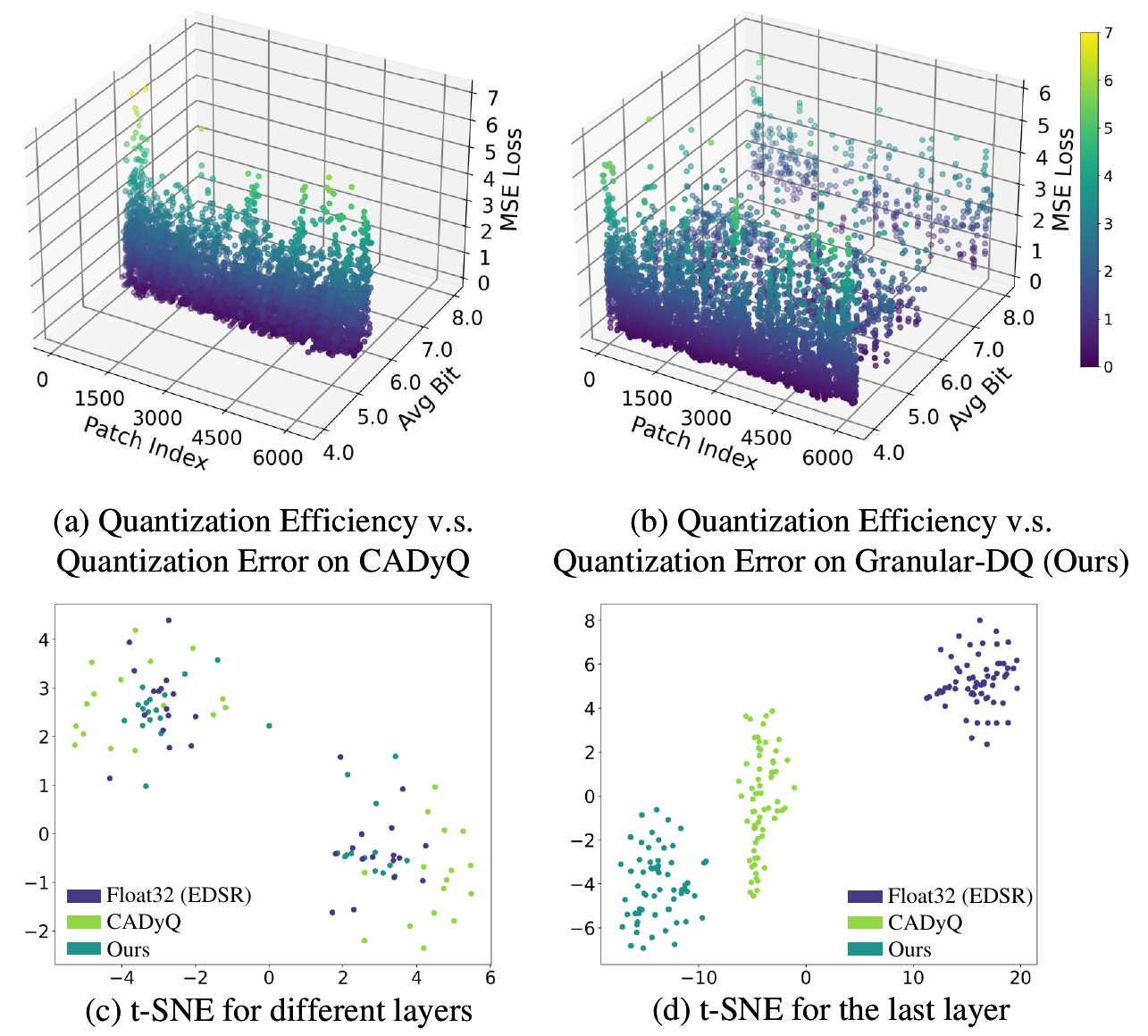}
	\caption{Analysis of the quantization efficiency, quantization error, and feature distribution in t-SNE on CADyQ and our Granular-DQ. (a) and (b) illustrate the quantization efficiency v.s. quantization error trade-off; (c) and (d) visualize the feature distribution of two resultant models and compare with the corresponding original one (Float32: EDSR).}
	\label{fig:motivation}
\end{figure}

Based on the above analysis, this work aims to design a dynamic quantization approach for diverse image contents while maintaining the representation ability of the original model. To this end, we rethink the image characteristics related to image quality from the granularity and information density. As we know, the fine-granularity representations reveal the texture complexity of local regions, while coarse ones express structural semantics of the overall scene. Besides, according to Shannon's Second Theorem~\cite{shannon}, the entropy statistic reflects the average information density and the complexity of pixel distributions given patches, which is directly correlated to the image quality. Therefore, we propose Granular-DQ, a markedly
different method that fully explores the granularity and entropy statistic of images to quantization adaption. Granular-DQ contains two sequential steps: 1) granularity-aware bit allocation for all the patches and 2) entropy-based fine-grained bit-width adaption for the patches less quantized by 1). In this way, we can see that the bit-width allocation by Graular-DQ is sparser than CADyQ, where a majority of patches are lower than 5bit with only a few patches at high bit-width (Figure~\ref{fig:motivation} (b)). Moreover, the feature distribution of the layers quantized by our method is closer to that of the original model (Figure~\ref{fig:motivation}(c)-(d)). These validate that our Granular-DQ enables low-bit and layer-invariant quantization. 

\renewcommand{\thetable}{S\arabic{table}}
\setcounter{table}{0} 
\begin{table}[t]
\footnotesize
\centering
\setlength\tabcolsep{4pt}
    \begin{tabularx}{\hsize}{@{\extracolsep{\fill}}llYYYYYY}
        \toprule
        \multirow{2}{*}{$t_1$} & \multirow{2}{*}{$t_2$} & \multicolumn{3}{c}{Set14} & \multicolumn{3}{c}{Urban100}\\
        \cmidrule(lr){3-5}  \cmidrule(lr){6-8} 
         & & FAB & PSNR & SSIM & FAB & PSNR & SSIM   \\ 
         \midrule
            0.4 & 0.7 & 5.86 & 28.53 & 0.780 & 5.28  & 25.96  &  0.783    \\ 
            0.4 & 0.8 & 5.86 & 28.50 & 0.779 & 5.04  & 25.99  &  0.783    \\
            0.4 & 0.9 & \textbf{5.54} & 28.56 & 0.780 & 4.99  & 25.98  &  0.782    \\   
        \midrule
            0.5 & 0.7 & 5.86 & 28.53 & 0.780 & 5.25  & 25.97  & \textbf{0.784}    \\
            0.5 & 0.8 & 5.82 & 28.57 & 0.780 & 5.02  & 26.00  & 0.782     \\
            0.5 & 0.9 & \textbf{5.54} & \textbf{28.58} & \textbf{0.781} & \textbf{4.97}  & \textbf{26.01}  & \textbf{0.784}   \\
        \hline   
\end{tabularx}
\caption{Ablation study on the impact of the thresholds in ATC with EDSR baseline.}
\label{sup_tab5}
\end{table}

\begin{table}[t]
\footnotesize
\centering
\setlength\tabcolsep{3pt}
    \begin{tabularx}{\hsize}{@{\extracolsep{\fill}}llYYYYYY}
        \toprule
        \multirow{2}{*}{$t^{*}$} & \multirow{2}{*}{$b^{*}$} & \multicolumn{3}{c}{Set14} & \multicolumn{3}{c}{Urban100} \\
        \cmidrule(lr){3-5} \cmidrule(lr){6-8} 
         &  & FAB & PSNR & SSIM &  FAB & PSNR & SSIM    \\ 
         \midrule
            $[0.5]$         &$[4,8]$     & 6.57  & 28.57 & 0.780 & 5.75 & 25.97 & 0.781     \\ 
            $[0.5,0.9]$     &$[4,5,8]$   &\textbf{5.54} & \textbf{28.58} &\textbf{0.781} & \textbf{4.97}  & \textbf{26.01} & \textbf{0.784}\\ 
            $[0.4,0.6,0.9]$ &$[4,5,6,8]$ & 6.07  & \textbf{28.58} & 0.779 & 5.41  & 25.93 & 0.781   \\
            $[0.4,0.6,0.9]$ &$[4,5,7,8]$ & 6.21  & 28.54 & 0.780 & 5.61  & 25.93 & 0.782   \\
        \hline   
\end{tabularx}
\caption{Ablation study on the influence of a different number of thresholds (quantile, denoted by $t^{*}$) and corresponding bit configuration (denoted by $b^{*}$) in E2B with EDSR.}
\label{sup_tab4}
\end{table}

\subsection{Ablation Study}
\noindent\textbf{Impact of the Threshold $t$ in ATC.} In this work, we set two thresholds $t_1$ and $t_2$ in ATC, which divide the entropy of input patches into 3 subintervals and then map them to the bit codes ([4, 5, 8] in Table~\ref{sup_tab5}), whitch facilitates the bit-width adjustment in E2B. As in Table~\ref{sup_tab5}, according to the results on Set14 and Urban100, we can empirically set the combination of [$t_1=0.5$, $t_2=0.9$] as it achieves the best balance in quantization. 

\noindent\textbf{Impact of the Threshold Number in E2B.} We further experimentally investigated the effect of different numbers of thresholds in E2B and their corresponding candidate bit configuration. Firstly, we assume that there is only one quantile for all the input patches, which means the entropy statistic $\mathbf{H}$ is divided into two subintervals. As shown in Table~\ref{sup_tab4}, when we adjust the bit-widths of patches using 4/8bit, the model performs worst on both Set14 and Urban100 datasets. Similarly, when we incorporate three thresholds of $t$ with $[0.4,0.6,0.9]$ to divide $\mathbf{H}$ into four subintervals, it can be seen that whether using the bit configurations of $[4, 5, 6, 8]$ or $[4, 5, 7, 8]$, the model cannot obtain satisfied quantization efficiency. In contrast, the model with two thresholds $[0.5,0.9]$ and corresponding candidate bit-widths of $[4, 5, 8]$ achieved the best trade-off on both datasets, making it our final choice.

\noindent\textbf{Influence of Different Quantization Patterns.} To investigate the compatibility of our method on different quantization patterns, we conduct experiments by combining Granular-DQ with PAMS~\cite{li2020pams} and QuantSR~\cite{qin2024quantsr}, where the results on Urban100 are reported in Table~\ref{sup_tab3}. Notably, different from existing dynamic methods~\cite{hong2022cadyq,Tian_2023_CVPR}, our Granular-DQ does not require the pre-trained models of PAMS or QuantSR. We can see that Granular-DQ+PAMS gets 0.07dB PSNR gains with 0.4 FAB reduction for EDSR compared to CADyQ+PAMS. When applying the QuantSR scheme on Granular-DQ, the model can achieve the best trade-off between FAB and PSNR/SSIM for both EDSR and IDN models, where even the latter surpasses the original model by 0.26dB in PSNR.

\begin{table}[t]
    \setlength\tabcolsep{3pt}
    \footnotesize
    \centering
    \begin{tabularx}{\hsize}{@{\extracolsep{\fill}}lccc}
        \toprule
        \multirow{2}{*}{\centering Methods}  & \multicolumn{3}{c}{Urban100}  \\ 
        \cmidrule(lr){2-4} 
         & FAB↓ & PSNR↑ & SSIM↑  \\
        \midrule
        EDSR   & 32.00 & 26.03  & 0.784  \\
        PAMS   & 8.00  & 26.01  & 0.784  \\
        CADyQ+PAMS  & 6.09  & 25.94  & 0.782   \\
        Granular-DQ+PAMS  &  5.69  & 25.95  & 0.782    \\
        Granular-DQ+QuantSR &  \textbf{4.97}  & \textbf{26.01}  & \textbf{0.784}    \\
        \midrule 
        IDN   & 32.00 & 25.42  & 0.763    \\
        PAMS  & 8.00  & 25.56  & 0.768    \\
        CADyQ+PAMS & 5.78  & 25.65  & 0.771     \\
        Granular-DQ+PAMS  & 4.73  & 25.62  & 0.770 \\
        Granular-DQ+QuantSR & \textbf{4.18}  & \textbf{25.68}  & \textbf{0.772} \\ 
        \midrule
    \end{tabularx}
    \caption{Investigation of the compatibility of our Granular-DQ with different quantization patterns. We observe the $\times 4$ SR results on Urban100 based on EDSR and IDN. }
    \label{sup_tab3}
\end{table}

\subsection{Implementation Details on Transformer-based Baselines}
For the transformer-based models, the linear layers of the MLPs in both SwinIR-light~\cite{liang2021swinir} and HAT-S~\cite{chen2023hat} are all quantized using the QuantSR scheme~\cite{qin2024quantsr}. Surprisingly, despite all our efforts, we still encountered the gradient explosion issue when implementing the quantization scheme for CADyQ~\cite{hong2022cadyq} and CABM~\cite{Tian_2023_CVPR} in HAT-S. As a result, these two retained full precision for channel attention in the experiments. During the training phase, we randomly cropped the LR image into 64 $\times$ 64 with a total batch size of 16 for all scale factors, following the settings of the original model. The learning rate was initially set to $2\times 10^{-4}$ and halved after 250K iterations.

\begin{table*}[t]
    \setlength\tabcolsep{3pt}
    \setlength{\abovecaptionskip}{5pt}
    \centering
    \small
    \begin{tabularx}{\textwidth}{@{\extracolsep{\fill}}llYYYYYYYYY}
        \toprule
        \multirow{2}{*}{\centering Methods} & \multirow{2}{*}{\centering Scale} & \multicolumn{3}{c}{Urban100} & \multicolumn{3}{c}{Test2K} & \multicolumn{3}{c}{Test4K} \\ 
        \cmidrule(lr){3-5} \cmidrule(lr){6-8} \cmidrule(lr){9-11} 
         & & FAB↓ & PSNR↑ & SSIM↑ & FAB↓ & PSNR↑ & SSIM↑ & FAB↓ & PSNR↑ & SSIM↑ \\
        \midrule
        SRResNet & $\times2$ & 32.00 & 32.11  & 0.928 & 32.00 & 32.81  & 0.930& 32.00 & 34.53  & 0.944   \\
        PAMS   & $\times2$ & 8.00  & 31.96  & 0.927 & 8.00  & 32.72  & 0.928& 8.00  & 34.33  & 0.943   \\
        CADyQ  & $\times2$ & 6.46  & 31.58  & 0.923 & 6.10  & 32.61  & 0.926 & 6.02  & 34.19  & 0.942   \\
        CABM   & $\times2$ & 5.46  & 31.54  & 0.923 & 5.33  & 32.55  & 0.925 & 5.23  & 34.16  & 0.942   \\ 
        Granular-DQ (Ours) & $\times2$ & \textbf{4.11}  & \textbf{31.94}  & \textbf{0.927}  & \textbf{4.17}  & \textbf{32.52}  & \textbf{0.925}  & \textbf{4.12}  & \textbf{34.52}  & \textbf{0.944}   \\
        \midrule
        EDSR   & $\times2$ & 32.00 & 31.97  & 0.927 & 32.00  & 32.75 & 0.928  & 32.00  & 34.38  & 0.943   \\
        PAMS   & $\times2$ & 8.00  & 31.96  & 0.927 & 8.00  & 32.72  & 0.928  & 8.00  & 34.33  & 0.943   \\ 
        CADyQ  & $\times2$ & 6.15  & 31.95  & 0.927 & 5.68  & 32.70  & 0.928  & 5.59  & 34.30  & 0.943   \\ 
        CABM   & $\times2$ & 5.59  & 31.92  & 0.927 & 5.39  & 32.74  & 0.927  & 5.31  & 34.33  & 0.943   \\ 
        Granular-DQ (Ours) & $\times2$ & \textbf{4.60}  & \textbf{32.01}  & \textbf{0.928}  & \textbf{4.40}  & \textbf{32.57}  & \textbf{0.925} & \textbf{4.27}  & \textbf{34.42} & \textbf{0.944}   \\
        \hline
        IDN   & $\times2$ & 32.00 & 31.29  & 0.920  & 32.00   & 32.42  & 0.924& 32.00 & 34.02  & 0.940   \\
        PAMS  & $\times2$ & 8.00  & 31.39  & 0.921 & 8.00   & 32.46  & 0.925  & 8.00  & 34.05  & 0.941   \\
        CADyQ & $\times2$ & 5.22  & 31.54  & 0.923 & 4.67   & 32.51  & 0.925  & 4.57  & 34.10  & 0.941   \\
        CABM  & $\times2$ & 4.21  & 31.40  & 0.921 & 4.19   & 32.50  & 0.925  & 4.19  & 34.10  & 0.941   \\
        Granular-DQ (Ours) & $\times2$ & \textbf{4.01}  & \textbf{31.63}  & \textbf{0.924} & \textbf{4.05}   & \textbf{32.36}  & \textbf{0.922} & \textbf{4.05}  & \textbf{34.35}  & \textbf{0.942}   \\ 
        \midrule  
        SwinIR-light & $\times2$ & 32.00 & 32.71  & 0.934   & 32.00 & 32.81  & 0.928   & 32.00 & 34.81  & 0.946   \\
        PAMS      & $\times2$ & 8.00  & 32.40  & 0.931   & 8.00  & 32.68  & 0.927   & 8.00  & 34.68  & 0.945   \\ 
        CADyQ     & $\times2$ & 5.29  & 31.88  & 0.926   & 5.07  & 32.50  & 0.924   & 5.06  & 34.48  & 0.943   \\ 
        CABM      & $\times2$ & 5.14  & 31.93  & 0.927   & 4.98  & 32.52  & 0.925   & 4.97  & 34.50  & 0.944   \\ 
        Granular-DQ (Ours)  & $\times2$ & \textbf{4.76}  & \textbf{32.54}  & \textbf{0.932} & \textbf{4.73}  & \textbf{32.73}  & \textbf{0.927} & \textbf{4.12}  & \textbf{34.52}  & \textbf{0.944}   \\ 
        \midrule
        HAT-S & $\times2$ & 32.00  & 34.19  & 0.945   & 32.00  & 33.28  & 0.934   & 32.00 & 35.30  & 0.950   \\
        PAMS      & $\times2$ & 8.00  & 33.63  & 0.941   & 8.00  & 33.12  & 0.932  & 8.00  & 35.12  & 0.949   \\ 
        CADyQ     & $\times2$ & 5.43  & 33.13  & 0.938   & 5.32  & 32.95  & 0.930  & 5.22  & 34.95  & 0.947   \\ 
        CABM      & $\times2$ & 5.34  & 33.09  & 0.937   & 5.26  & 32.94  & 0.930  & 5.18  & 34.95  & 0.947 \\  
        Granular-DQ (Ours)  & $\times2$ & \textbf{4.80}  & \textbf{33.71}  & \textbf{0.942}  & \textbf{4.78} & \textbf{33.12}  & \textbf{0.932}  & \textbf{4.77}  & \textbf{35.12}  & \textbf{0.949}   \\
        \midrule
    \end{tabularx}
    \caption{Quantitative comparison (FAB, PSNR (dB)/SSIM) with full precision models, PAMS, CADyQ, CABM and our method on Urban100, Test2K, Test4K for $\times2$. }
    \label{sup_tab1}
\end{table*}

\subsection{Comparison with the State-of-the-Art}
\noindent\textbf{Quantitative Comparison for $\times$2 SR.} We further conduct experiments for $\times 2$ SR, where the quantitative results are illustrated in Table~\ref{sup_tab1}. Obviously, Granular-DQ demonstrates competitive trade-offs in terms of FAB and PSNR/SSIM compared to other quantization methods across all CNN models. Additionally, for SwinIR-light and HAT-S, Granular-DQ also achieves remarkably superior reconstruction accuracy to full precision than others while maintaining the lowest FAB.

\begin{figure*}[t]
	\centering
	\includegraphics[width=\linewidth]{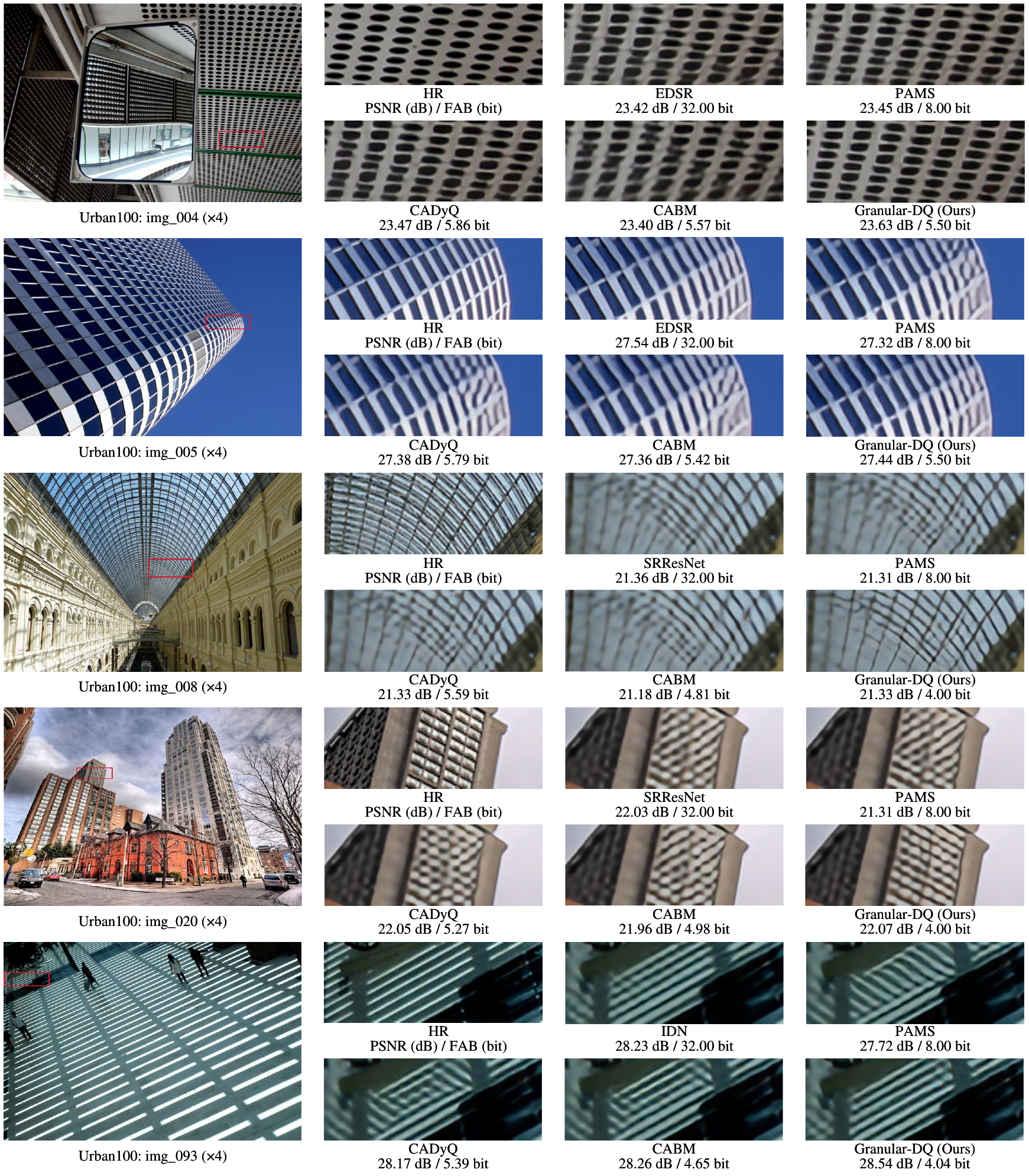}
	\caption{More visual comparison ($\times$4) on Urban100 ($\times4$) for different methods.}
	\label{fig:sup_sr1}
\end{figure*}

\begin{figure*}[t]
	\centering
	\includegraphics[width=\linewidth]{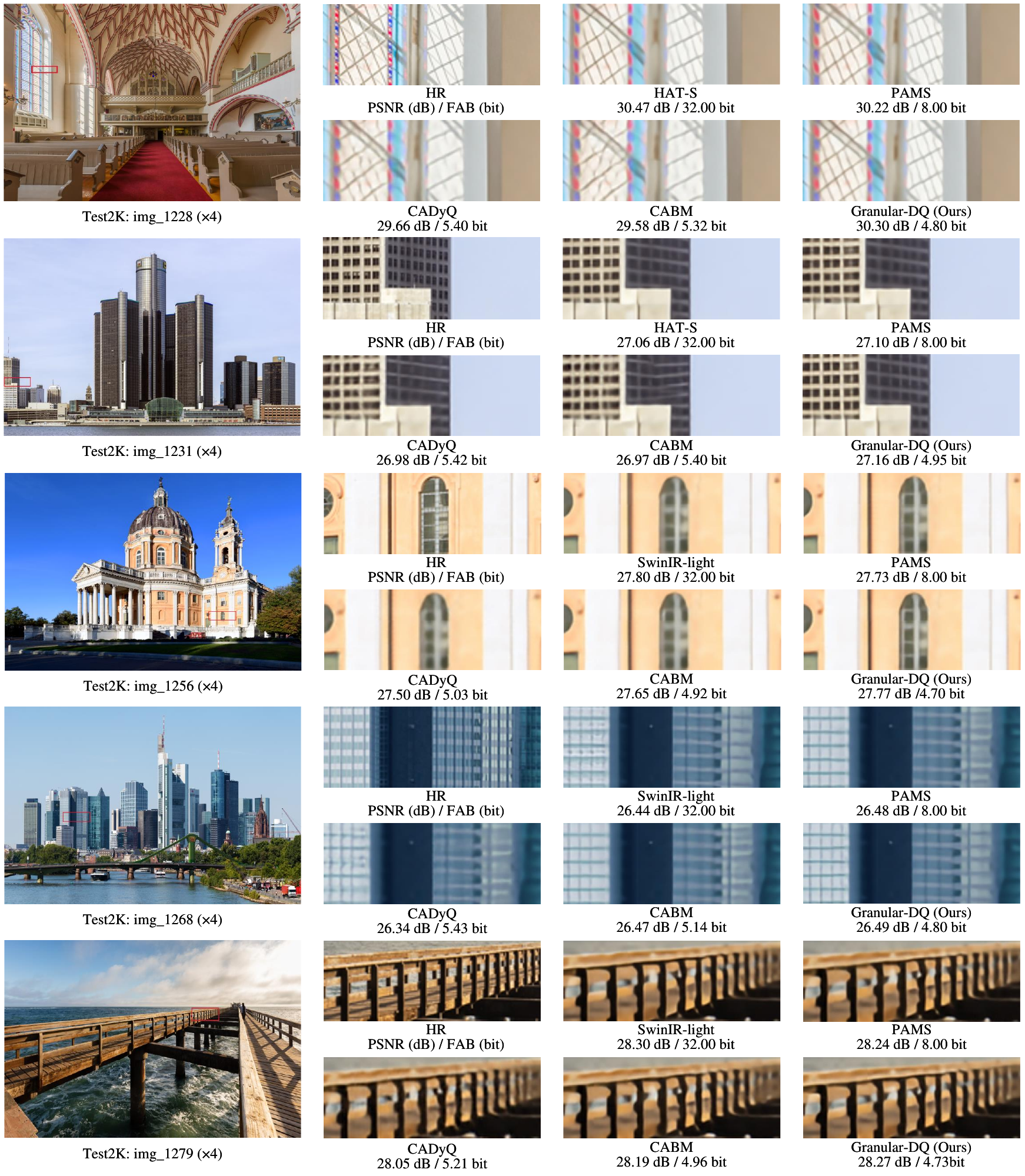}
	\caption{More visual comparison ($\times$4) on Test2K ($\times4$) for different methods.}
	\label{fig:sup_sr2}
\end{figure*}

\noindent\textbf{More Qualitative Comparison.} 
In Figure~\ref{fig:sup_sr1} and~\ref{fig:sup_sr2}, we provide more $\times 4$ SR visual results produced by recent state-of-the-art methods and our Granular-DQ. Regardless of whether the models are CNN-based or transformer-based, Granular-DQ consistently achieves superior reconstruction details at the lowest FAB compared to other quantization methods in most instances. In each case, minimal discrepancies can be observed between Granular-DQ and its corresponding full-precision model. These findings further validate that Granular-DQ ensures an optimal trade-off between reconstruction accuracy and quantization efficiency.  

\subsection{Limitation}
 While Granular-DQ effectively maintains promising SR performance with dramatic computational overhead reduction, it still has several limitations. First, the mixed-precision solution of Granular-DQ makes it require specific hardware design and operator support to achieve true compression acceleration. Second, its efficacy in accelerating processing for super-resolving large-size images is modest at best. In future work, we will design more efficient and effective quantization approaches to overcome these limitations.


\end{document}